\documentclass[sigconf,natbib]{acmart}

\usepackage{sigir24-generative-ir-evaluation-frame}
\graphicspath{{../sigir24-generative-ir-evaluation-figures/}}

\begin{document}
\title{Evaluating Generative Ad Hoc Information Retrieval}

\author[L.\ Gienapp]{Lukas Gienapp}
\authornotemark[1]
\affiliation{
\institution{Leipzig University and ScaDS.AI}
\city{Leipzig}
\country{Germany}
}

\author[H.\ Scells]{Harrisen Scells}
\affiliation{
\institution{Leipzig University}
\city{Leipzig}
\country{Germany}
}

\author[N.\ Deckers]{Niklas Deckers}
\affiliation{
\institution{Leipzig University and ScaDS.AI}
\city{Leipzig}
\country{Germany}
}

\author[J.\ Bevendorff]{Janek Bevendorff}
\affiliation{
\institution{Leipzig University}
\city{Leipzig}
\country{Germany}
}

\author[S.\ Wang]{Shuai Wang}
\affiliation{
\institution{The University of Queensland}
\city{Brisbane}
\country{Australia}
}

\author[J.\ Kiesel]{Johannes Kiesel}
\affiliation{
\institution{Bauhaus-Universit{\"a}t Weimar}
\city{Weimar}
\country{Germany}
}

\author[S.\ Syed]{Shahbaz Syed}
\affiliation{
\institution{Leipzig University}
\city{Leipzig}
\country{Germany}
} 

\author[M.\ Fr{\"o}be]{Maik Fr{\"o}be}
\affiliation{
\institution{Friedrich-Schiller-Universit{\"a}t Jena}
\city{Jena}
\country{Germany}
}

\author[G.\ Zuccon]{Guido Zuccon}
\affiliation{
\institution{The University of Queensland}
\city{Brisbane}
\country{Australia}
}

\author[B.\ Stein]{Benno Stein}
\affiliation{
\institution{Bauhaus-Universit{\"a}t Weimar}
\city{Weimar}
\country{Germany}
}

\author[M.\ Hagen]{Matthias Hagen}
\affiliation{
\institution{Friedrich-Schiller-Universit{\"a}t Jena}
\city{Jena}
\country{Germany}
}

\author[M.\ Potthast]{Martin Potthast}
\affiliation{
\institution{\mbox{\kern-1em University of Kassel, hessian.AI, ScaDS.AI}}
\city{Kassel}
\country{Germany}
}

\authornote{
Corr. Auth. \texttt{lukas.gienapp@uni-leipzig.de} / \texttt{martin.potthast@uni-kassel.de}
}
\renewcommand{\shortauthors}{Lukas Gienapp et al.}

\begin{abstract}
Recent advances in large language models have enabled the development of viable generative retrieval systems. Instead of a traditional document ranking, generative retrieval systems often directly return a grounded generated text as a response to a query. Quantifying the utility of the textual responses is essential for appropriately evaluating such generative ad hoc retrieval. Yet, the established evaluation methodology for ranking-based ad hoc retrieval is not suited for the reliable and reproducible evaluation of generated responses. To lay a foundation for developing new evaluation methods for generative retrieval systems, we survey the relevant literature from the fields of information retrieval and natural language processing, identify search tasks and system architectures in generative retrieval, develop a new user model, and study its operationalization.
\end{abstract}




\copyrightyear{2024}
\acmYear{2024}
\setcopyright{rightsretained}
\acmConference[SIGIR '24]{Proceedings of the 47th International ACM SIGIR Conference on Research and Development in Information Retrieval}{July 14--18, 2024}{Washington, DC, USA}
\acmBooktitle{Proceedings of the 47th International ACM SIGIR Conference on Research and Development in Information Retrieval (SIGIR '24), July 14--18, 2024, Washington, DC, USA}
\acmDOI{10.1145/3626772.3657849}
\acmISBN{979-8-4007-0431-4/24/07}


\makeatletter
\gdef\@copyrightpermission{
 \begin{minipage}{0.3\columnwidth}
  \href{https://creativecommons.org/licenses/by/4.0/}{\includegraphics[width=0.90\textwidth]{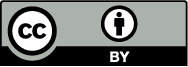}}
 \end{minipage}\hfill
 \begin{minipage}{0.7\columnwidth}
  \href{https://creativecommons.org/licenses/by/4.0/}{This work is licensed under a Creative Commons Attribution International 4.0 License.}
 \end{minipage}
 \vspace{5pt}
}
\makeatother

\maketitle

\section{Introduction}

\begin{figure}[t]
\includegraphics[width=\linewidth]{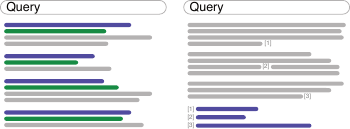}
\caption{A search engine results page~(SERP) has traditionally been a list of document references (list~SERP, left). Many generative retrieval systems now have ``reinvented'' SERPs as generated texts with references (text~SERP, right).}
\label{list-serp-text-serp}
\Description{On the left, a schematic illustration of a list of search results as it is shown on conventional search results pages. On the right, a schematic text with citations and a reference list of corresponding links below it as a novel presentation of a search result.}
\end{figure}

The development of large language models~(LLMs) has prompted search engines to innovate the way results are presented: using LLMs to directly generate a textual response from a query's results. While LLMs can generate unreliable information~\cite{alkaissi:2023,ji:2023,koopman:2023b}, conditioning their inference o n relevant search results has emerged as a potential technique to ground generated statements~\cite{lewis:2020,mialon:2023}. As textual answers can relieve users of the (cognitive) effort of collecting the needed information from individual search results themselves, the design of some search engine's results pages~(SERPs) has changed (Figure~\ref{list-serp-text-serp}): instead of the proverbial list of ``ten blue links'' (list~SERP, left), a generated text with references is shown (text~SERP, right). The first public prototypes of this kind were You.com's You~Chat and Neeva~AI, closely followed by Microsoft's Bing Copilot, Google's Gemini, Perplexity.ai, Baidu's Ernie,%
\footnote{\raggedright See \url{https://chat.you.com}; Neeva has shut down; \url{https://chat.bing.com}; \url{https://gemini.google.com}; \url{https://perplexity.ai}; \url{https://yiyan.baidu.com}.}
and other research prototypes~\cite{koopman:2023,zhang:2023}.
Far ahead of this development, already in~\citeyear{sakai:2011} \citet{sakai:2011} raised an important question: how can text~SERP-based search engines be evaluated? An answer was and is not that straightforward, since the modern theory and practice of retrieval evaluation is premised on the assumption that search results are presented as list~SERPs.%
\footnote{Even though research on search interfaces has suggested and studied many interaction designs and variants of result presentation~\cite{hearst:2009, wilson:2011, liu:2021}, with the growth of the Web, the list~SERP design became a de facto standard for web search.}

According to list~SERP user models, a ranked list of results triggers a certain user behavior like reading the results in order until the information need is satisfied or the search is abandoned. In decades of research, a comprehensive theoretical framework of reliable and validated evaluation methods has been built to assess the quality of result rankings with respect to information needs. Replacing ranked results by a generated text undermines this foundation.

In this paper, we focus on questions related to transferring established list~SERP evaluation methodology to text~SERPs. Our approach is theory-driven and based on a systematic analysis of relevant literature from information retrieval~(IR) and related fields. Our contributions relate to the system, user, and evaluation perspectives. Starting with a definition of what generative ad hoc retrieval is, we distinguish two fundamental system models for generative retrieval and contextualize them in \citeauthor{broder:2002}'s~\cite{broder:2002} taxonomy of search tasks~(Sec\-tion~\ref{background-and-related-work}). We then devise a user model for text~SERPs, grounded in related behavioral studies (Section~\ref{user-model-for-generative-ir}). Finally, we revisit IR~evaluation methodologies to develop a foundation for text~SERP effectiveness measures and for the reliable evaluation of generative ad hoc retrieval (Section~\ref{operationalizing-evaluation}).

\section{Generative Retrieval}
\label{background-and-related-work}

In this section, we define the task of generative ad hoc retrieval, we review the two fundamental paradigms of its operationalization, we discuss its contribution on top of traditional ad hoc retrieval, and we distinguish it from other generative retrieval tasks. 

\subsection{Generative Ad Hoc Retrieval}

\begin{figure}[t]
\centering
\begin{tikzpicture}[
    every node/.style = {
        font=\normalsize\sffamily, 
        text=Black,
        text width=1.35cm,
        minimum height=.5cm,
        align=center,
        inner sep=0
    },
    every label/.style = {
        font=\small\sffamily, 
        text=Gray5,
        text width=2.1cm,
        inner sep=0
    },
    Label/.style = {
        font=\small\sffamily, 
        text=Gray5,
        text width=2.1cm
    },
    every path/.style = {
        -stealth,
        line width=.8pt,
        color=Gray6
    }
]

\node[label={[text width=3cm, label distance=.1cm]90:All possible texts}] (anchor) {$\mathcal{T}$};
\node[left = -.2cm of anchor, label={[label distance=.25cm]180:Queries / prompts}] (queries) {$Q$};
\node[right = -.2cm of anchor, label={[label distance=.25cm]0:Documents}] (documents) {$D$};

\node[below = 1.25cm of queries, label={[label distance=.25cm]180:Retrieval model}] (retrieve) {$\rho$};
\node[below = 1.25cm of documents, label={[label distance=.25cm]0:Language model}] (generate) {$\psi$};

\node[text=Red4, below = 1.25cm of retrieve, label={[label distance=.25cm]180:Retrieval-then-generation}] (rag) {$\psi(\rho(q))$};
\node[text=Red4, below = 1.25cm of generate, label={[label distance=.25cm]0:Generation-then-retrieval}] (gar) {$\rho(\psi(q))$};

\path[] (anchor) -- node[pos=.5]{$\cong$} (queries);
\path[] (anchor) -- node[pos=.5]{$\cong$} (documents);

\draw (queries) -- node[pos=.5, left=.95cm, Label]{Indexing \& learning to rank} (retrieve);
\draw (queries) -- (generate);
\draw (documents) -- (retrieve);
\draw (documents) -- node[pos=.5, right=.95cm, Label]{Training \& alignment} (generate);
\draw (retrieve) -- node[pos=.5, left=.95cm, Label]{Combination} (rag);
\draw (generate) -- (rag);
\draw (retrieve) -- (gar); 
\draw (generate) -- node[pos=.5, right=.95cm, Label]{Combination} (gar);

\end{tikzpicture}
\caption{In generative ad hoc retrieval, a retrieval model is combined with a language model. The notation assumes~$\rho$ and $\psi$ have texts from $\mathcal{T}$ as input and output, and that they can be complex pieces of software, like Google or ChatGPT.}
\label{generative-retrieval-specification}
\Description{}
\end{figure}
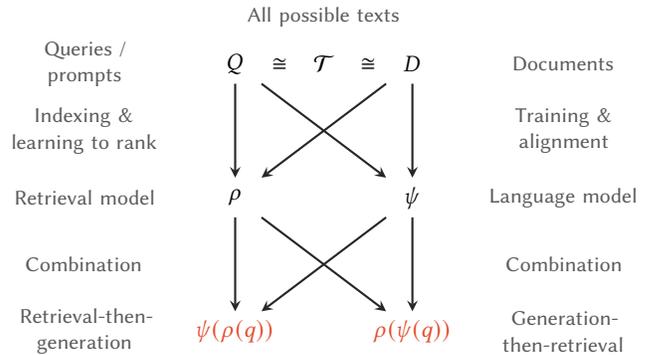

Ad hoc retrieval refers to scenarios where a user submits a single query and expects the underlying information need to be satisfied by a single result set (i.e., the information need must be satisfied without knowing any previous queries or interactions). At first glance, this ad hoc retrieval task and the task of language generation seem to be quite different. However, retrieval systems and generative language models are both built using document collections~$D$ (see Figure~\ref{generative-retrieval-specification}, top), and the usefulness of both depends on tuning them with user needs, expressed as queries or prompts~$Q$. Users of a retrieval system want to retrieve the most relevant documents for a query, and users of a generative language model want it to generate the most helpful text for a prompt. From an IR~perspective, the most salient difference is that a retrieval model~$\rho$ induces a ranking on a \emph{finite} document collection~$D$, while a generative language model~$\psi$ induces a ranking on the \emph{infinite} set of all possible texts~$\mathcal{T}$; generative models were thus recently also framed as infinite indexes~\cite{deckers:2023}. In practice, though, retrieval models only return the top-$k$ results, and generative language models only return one of the possibly many relevant texts from~$\mathcal{T}$.

As a retrieval model~$\rho$ can only return existing documents, the information available in the underlying collection~$D$ determines the degree to which a user's information need can be satisfied. Still, the user has to examine the returned documents for the desired information. A generative language model~$\psi$ instead attempts to alleviate the effort of examining documents by returning a tailored response that integrates all desired information. Yet, the factual accuracy of current generative language models is often prone to confabulations or hallucinations~\cite{zhao:2023,koopman:2023b,ji:2023,alkaissi:2023} (i.e., there is only a very small subset of accurate texts among all possible texts~$\mathcal{T}$).%
\footnote{For counterfactual information needs (e.g., What if Columbus didn't discover America?~\cite{kiciman:2023}), strong confabulation capabilities could be explicitly desirable, though.}

The term `generative ad hoc retrieval' refers to approaches that combine the advantages of retrieval and generation in ad hoc scenarios (one query, one result) by retrieving relevant documents from~$D$ and generating an answer from them, or by generating a response and ``verifying'' its statements by retrieving supporting documents from~$D$ (Figure~\ref{generative-retrieval-specification}, bottom left resp. right).

\subsection{Two Operationalization Paradigms}\label{paradigms}

Systems for generative ad hoc retrieval require a retrieval component to gather existing documents from a collection for a query, and a generation component to generate a text for a prompt. These components can be combined following two different paradigms~\cite{gao:2023}: \emph{retrieval-then-generation} or \emph{generation-then-retrieval} (Figure~\ref{generative-retrieval-specification}, bottom). In a retrieval-then-generation approach, a language model is conditioned with retrieved source material, for instance, by adding evidence to its input prompt~\cite{izacard:2021,khot:2023,lazaridou:2022,shi:2023}, by attending to retrieved sources during inference~\cite{lewis:2020,guu:2020,borgeaud:2022}, by chaining ideas~\cite{jiang:2022}, or by iterative self-attention~\cite{zhang:2021b}. In a generation-then-retrieval approach, the retrieval model is used to find sources for generated text passages. Though this idea has received less attention so far~\cite{autogpt:2023}, it resembles retrieving references for individual generated statements, similar to claim verification~\cite{wadden:2020}.

With increasing inference speeds of generative language models, arbitrarily ordered combinations of multiple retrieval and generation steps are possible, leading to \emph{multi-step generative ad hoc retrieval}. The simplest form might be iterative cycles like generating a text passage that is used as a query to retrieve relevant sources, which in turn serve as context for the next generation, etc. Applications are the continuous generation of text~\cite{ram:2023,jiang:2023,semnani:2023}, retrieving sources in multiple steps~\cite{patwardhan:2023}, or the refinement of a text through iterative inference~\cite{khattab:2022,autogpt:2023}.

In this paper, we focus on the evaluation of the text~SERP output of (possibly multi-step) generative ad hoc retrieval, but we do not consider evaluating any step individually.

\subsection{Generative (Ad Hoc) Search Tasks}
\label{generative-search-task}

In~2002, \citeauthor{broder:2002} suggested a now well-known taxonomy of search tasks~\cite{broder:2002} and related them to three generations of web search systems (see Table~\ref{table-search-task-overview}). Each generation utilizes a new source of information in addition to those of its predecessors to meet new user intents. First-generation systems support informational tasks, relying only on the information found within some single document to support a user's intent to acquire (parts of) that information. Second-generation systems additionally exploit document relations, supporting users to reach a specific site, document, or the most authoritative one among many alternatives (i.e., navigational tasks). Third-generation systems blend results from different possibly multimodal vertical systems into a single~SERP to support a user in performing transactional tasks.

We argue that generative retrieval systems can be seen as a new fourth generation of web search systems. Their synthesis of a single result ``document'' that condenses information from different sources relevant to some information need promises to reduce the users' cognitive load compared to prior system generations that required users to condense the information themselves.%
\footnote{\citet{sakai:2011} had proposed to present lists of short automatically identified relevant information nuggets instead of complete documents in \citeyear{sakai:2011}, but they had not considered the aspect of condensing the nuggets to a single result.}
Additionally, the ``synthesizing'' nature of generative retrieval systems can conceivably be exploited to generate new pieces of information not contained in the retrieved sources, rendering the generative model itself another new source of information.

\begin{table}
\centering
\caption{Top rows: \citeauthor{broder:2002}'s~(\citeyear{broder:2002}) identified generations of web search systems~(Gen.) and the tasks from his taxonomy~\cite{broder:2002} that each generation additionally supports~(+). Bottom row: Generative retrieval systems constitute a new 4th~generation that aids users in ``synthetic'' search tasks that require a system to synthesize and condense information.}
\renewcommand{\tabcolsep}{3pt}
\renewcommand{\arraystretch}{.8}
\begin{tabular}{@{}llllc@{}}
\toprule
  \bf Gen.              & \bf Search task & \bf Information source & \bf User intent & \bf Year \\
\midrule
  1\textsuperscript{st} & informational   & Document               & Acquire         & 1995 \\
  2\textsuperscript{nd} & + navigational  & + Document relations   & + Reach         & 1998 \\
  3\textsuperscript{rd} & + transactional & + Search verticals     & + Perform       & 2002 \\
\midrule
  4\textsuperscript{th} & + synthetic   & + Generative models    & + Condense      & 2023 \\
\bottomrule
\end{tabular}

\label{table-search-task-overview}
\end{table}

While many of the search tasks addressed by generative retrieval systems may seem to be informational in nature, we still suggest to also separate the search tasks in a new category of \emph{synthetic search tasks}. Complex needs like argumentative questions (Should society invest in renewable energy?) or decision-making questions (Should I get life insurance?) are simply not represented that well in \citeauthor{broder:2002}'s original categories. In contrast to informational tasks, the required information is hardly contained in some single document but rather spread across multiple documents; in contrast to navigational tasks, no single page is anticipated by the user to be reached; and in contrast to transactional tasks, the information condensation should be performed on the retrieval system side but not on the user side. Interestingly, as if already foreseeing generative retrieval, \citeauthor{broder:2002} even explicitly constrained informational queries and first-generation systems to static content: ``The purpose of such [informational] queries is to find information assumed to be available on the Web in a \emph{static form}. No further interaction is predicted, except reading. By static form we mean that the \emph{target document is not created} in response to the user query.''~\cite[page~5]{broder:2002}.

The new fourth generation of web search systems supports synthetic search tasks and enables users to access a single, comprehensive generated answer document that can cover in-depth analyses of multiple perspectives on some complex information need. Although the Web may actually also offer some set of documents to satisfy complex needs, an ideal generative retrieval system can directly \textit{dynamically} address them by retrieving relevant documents, synthesizing missing information, and condensing a coherent answer grounded in the retrieved sources.

\subsection{A Taxonomy of Generative Retrieval}

\begin{figure}[t]
\includegraphics{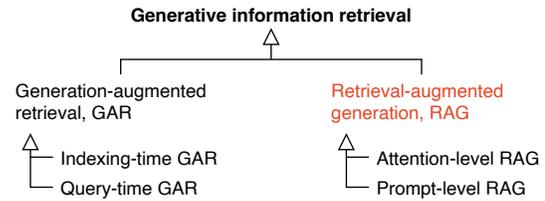}
\caption{Taxonomy of generative information retrieval and its two main instantiations: gen\-er\-a\-tion-augmented retrieval~(GAR, yielding list SERPs) and retrieval-augmented generation~(RAG, yielding text SERPs; focus of this paper).}
\Description{
A taxonomy graph that has generative information retrieval at the top level, and then is subdivided into generation-augmented retrieval (GAR) and retrieval-augmented generation (RAG) on the second level. The GAR node further consists of indexing-time GAR and query-time GAR, while the RAG node further consists of attention-level RAG and prompt-level RAG.
}
\label{generative-retrieval-taxonomy}
\end{figure}

`Generative retrieval' or `generative~IR' are umbrella terms for a diversity of approaches that use generative models to solve retrieval tasks.%
\footnote{See also the recent SIGIR~workshop on generative IR~\cite{benedict:2023}.}
Following \citet{arora:2023}, Figure~\ref{generative-retrieval-taxonomy} categorizes these approaches into generation-augmented retrieval~(GAR) and retrieval-augmented generation~(RAG). Notably, GAR~approaches create traditional list~SERPs, while RAG~approaches generate text~SERPs.

In GAR~approaches, generative models are used to enhance the traditional search architecture at indexing time or at query time. At indexing time, generative models can be used for augmenting documents~\cite{nogueira:2019,gospodinov:2023,macavaney:2020,formal:2021,zhuang:2021} with confabulated or hallucinated content, or for replacing the standard indexing process with what are commonly termed `differentiable indices' by, for instance, generating document identifiers like page titles~\cite{decao:2021,thorne:2022,chen:2022}, URLs~\cite{ziems:2023}, or (structured) string identifiers~\cite{zhou:2023,tay:2022,zhuang:2022,wang:2022}.
At query time, generative models can be used for augmenting queries~\cite{macavaney:2021,alaofi:2023}, or for modeling relevance by, for instance, generating parts of existing documents from the query and retrieving the documents by string matching~\cite{bevilacqua:2022}, by predicting a \mbox{(re-)}ranking directly~\cite{sun:2023}, or by using special tokens as relevance signal~\citep{nogueira:2020,ma:2023,qin:2023,zhuang:2023}.

In RAG~approaches---the focus of our paper---, generative models are augmented with retrieval capabilities; either internally as `attention-level~RAG', where the context attended to during generation is retrieved concurrently~\cite{guu:2020,jiang:2022,borgeaud:2022}, or externally as `prompt-level~RAG', where the retrieved context is inserted into the prompt. Orthogonally, one can distinguish the RAG~variants retrieval-then-gen\-er\-a\-tion and generation-then-retrieval (cf.~Section~\ref{paradigms}). Beyond GAR and RAG, generative models can also be used to directly generate a response without relying on retrieved information~\cite{robinson:2023}, i.e., as infinite indexes~\cite{deckers:2023}. This may involve generating multiple candidates and selecting the best one or regenerating a new response conditioned on the previous ones~\cite{yu:2023}. Moreover, an answer to a generative ad hoc request can also be the first turn of a conversational search~\cite{salton:1969,radlinski:2017,culpepper:2018}, where generative models have led to new tools~\cite{zhang:2021a,miller:2017} and dialog options~\cite{zamani:2020}. 

\section{A User Model for Generative IR}
\label{user-model-for-generative-ir}

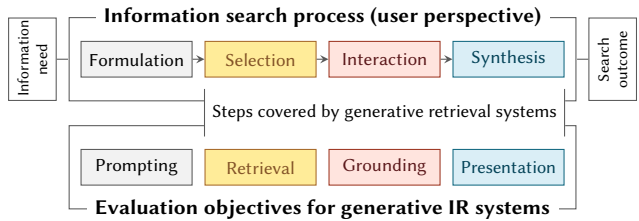
\begin{figure}
\begin{tikzpicture}
[
    every node/.style = {font=\footnotesize\sffamily,shape=rectangle},
    Boxed/.style = {draw=Gray5},
    Process/.style   = {minimum width = 4em, minimum height = 1.5em, text width=4em, align=center},
    Objective/.style = {minimum width = 4em, minimum height = 1.5em, text width=4em, align=center},
    Label/.style = {font=\small\sffamily\bfseries, fill=white, text=black},
    Annotation/.style = {font=\footnotesize\sffamily, text=black},
    Prompting/.style = {fill = Gray1, draw=Gray5, text=black},
    Retrieval/.style = {fill = Yellow1, draw=Yellow5, text=Yellow6},
    Grounding/.style = {fill = Red1, draw=Red5, text=Red6},
    Response/.style =  {fill = Blue1, draw=Blue5, text=Blue6},
    IO/.style =  {font=\scriptsize\sffamily,text width = 2.8em, align=center, draw=Gray5, minimum height=2em, text=black},
]

\coordinate (process anchor) at (-.36,1.5);
\coordinate (objectives anchor) at (0,0);

\begin{scope}[local bounding box=bbox1]
\node (formulate)  [Process, right=.5em of process anchor, anchor = west, Prompting] {Formulation};
\node (select)     [Process, right=.5em of formulate,      anchor = west, Retrieval] {Selection};
\node (interact)   [Process, right=.5em of select,         anchor = west, Grounding] {Interaction};
\node (synthesize) [Process, right=.5em of interact,       anchor = west, Response] {Synthesis};
\end{scope}
\draw[Boxed] ($(bbox1.south west) - (.5em,1em)$) rectangle ($(bbox1.north east) + (.5em,1em)$);
\node at ($(bbox1.north) + (0em,1em)$) [Label, anchor=center] {Information search process (user perspective)};
\node (input)  [IO, rotate=90, left=1em of formulate, anchor = south] {\kern-0.25em Information need};
\node (output)  [IO, rotate=90, right=1em of synthesize, anchor = north] {Search outcome};

\begin{scope}[local bounding box=bbox2]
\node (prompt)    [Objective, below= 3em of formulate,  anchor = north, Prompting] {Prompting};
\node (retrieval) [Objective, below= 3em of select,     anchor = north, Retrieval] {Retrieval};
\node (grounding) [Objective, below= 3em of interact,   anchor = north, Grounding] {Grounding};
\node (response)  [Objective, below= 3em of synthesize, anchor = north, Response]  {Presentation};
\end{scope}
\draw[Boxed] ($(bbox2.south west) - (.5em,1em)$) rectangle ($(bbox2.north east) + (.5em,1em)$);
\node at ($(bbox2.south) + (0em,-1em)$) [Label, anchor=center] {Evaluation objectives for generative IR systems};

\begin{scope}[local bounding box=bbox3]
\path ($(select.west) - (0em,1.25em)$) rectangle ($(synthesize.east) - (0em,3.25em)$) node [pos=0.5] (search system 1) {};
\end{scope}
\draw[fill=white, draw=white] (bbox3.south west) rectangle (bbox3.north east);
\draw[draw=white] (bbox3.south west) rectangle (bbox3.north east);
\path[draw=Gray5] (bbox3.north west) -- (bbox3.south west)
                  (bbox3.north east) -- (bbox3.south east);
\node (label) at (search system 1.center) [anchor=center, fill=white, text=black, inner sep=0] {Steps covered by generative retrieval systems};
 
\draw[-{stealth}, color=Gray5]
    (formulate)  edge (select)
    (select)     edge (interact)
    (interact)   edge (synthesize);

\draw[color=Gray5] 
    (input.south) -- ($(input.south) + (.15, 0)$)
    ($(output.north) - (.15, 0)$) -- (output.north);
  
\end{tikzpicture}
\caption{The information search process \cite{vakkari:2016} transforms an information need into a search outcome (top row). Respective corresponding evaluation objectives allow the derivation of a user model for an evaluation setting. Generative IR~systems cover the steps of `selection', `interaction', and `synthesis', for which we formulate the corresponding evaluation objectives `retrieval', `grounding', and `presentation' (bottom row).}
\Description{The figure consist of two aligned rows. The upper row illustrates the information search process from a searcher's perspective, with the four components of `formulation', `selection', `interaction', and `synthesis'. The lower row illustrates evaluation objectives for generative IR systems, with the four components `prompting', `retrieval', `grounding', and `presentation'. Three pairs of components are related: selection and retrieval, interaction and grounding, as well as synthesis and presentation. These three pairs represent the steps supported by generative IR systems.}
\label{fig:evaluation-overview}
\end{figure}

The general structure of an information search process~\cite{vakkari:2016} as seen from the users' perspective is shown in Figure~\ref{fig:evaluation-overview}~(top row). After formulating an information need and selecting and interacting with some search results, in a final synthesis step the users try to reach a satisfying outcome. While traditional list~SERPs mainly assist the users during selection and interaction, the text~SERPs of generative systems also directly encompass the synthesis step. 
Evaluating retrieval systems with respect to the information search process often relies on some model of user expectations and behavior. Yet, most current user models focus on list~SERPs but not text~SERPs. Thus, after preliminary considerations~(Sec\-tion~\ref{preliminary-considerations}), we explore how the information search process relates to generative approaches~(Sec\-tion~\ref{steps-of-search-process}). Afterwards, we follow the evaluation methodology proposed by~\citet{agosti:2014}: We define generative~IR-oriented evaluation objectives for the search process~(Sec\-tion~\ref{evaluation-objectives}; shown in the bottom row of Figure~\ref{fig:evaluation-overview}) and we devise a user model corresponding to these objectives~(Sec\-tion~\ref{components-of-user-model}). In Sec\-tion~\ref{operationalizing-evaluation}, we then operationalize the user model.

\subsection{Preliminary Considerations}
\label{preliminary-considerations}

\paragraph{Evaluation Setting}
Traditional search results (list~SERPs) are ranked lists of documents, each typically referenced by a linked title, snippet, and URL. In generative~IR, instead, the search result is a textual response (text~SERP), i.e., a sequence of statements, each optionally referenced to sources of evidence. A statement can be any consecutive passage of text, ranging from phrases to sentences or even longer paragraphs. In this context, we consider statements as atomic in the sense that we disregard the nesting of statements of different lengths, and in the sense that statements support claims that are pertinent to the user's information need---comparable to the concept of `atomic/semantic content units'~\cite{liu:2023b,nenkova:2007} in summarization evaluation, or `information nuggets'~/~`retrieval units' in tra\-di\-tio\-nal~IR~\cite{dang:2007,sakai:2011,sakai:2023,chen:2023}. A statement can be referenced to none, one, or more sources in form of explicit links to web documents containing the information on which the generated statement is based and by which it is grounded. In this paper, we consider the evaluation to be ad hoc, i.e., based on a single query without search session-based or conversational elements.

\paragraph{Evaluation Paradigms}
To estimate the effectiveness of list~SERP-based retrieval systems, offline evaluation following the Cranfield paradigm~\cite{cleverdon:1967} is a de facto standard in IR~research. The users' satisfaction with the results for a given topic~(query) is estimated by deriving effectiveness scores based on judging a pool of documents returned by the evaluated systems~\cite{sanderson:2010}. The pools often are also reused later to evaluate new search systems by checking whether their retrieved results previously were judged---and simplistically assuming non-relevance for the results without previous judgments~\cite{froebe:2023}. However, as the output ``documents'' of generative retrieval systems may be novel every time, simply assuming non-relevance would not lead to helpful evaluation results. Instead, more sophisticated transfer methods are required to adapt offline evaluation to generative retrieval.
Besides offline evaluation, generative retrieval systems could also be evaluated in an online fashion~\cite{sallam:2023}. Online evaluation does not rely on previous judgments but tries to estimate the output of some system by collecting explicit or implicit user feedback~\cite{kelly:2009} like user satisfaction ratings or clicks. This form of evaluation increases the manual effort, often happens in uncontrolled setups, may be expensive and time-consuming to conduct, and is challenging to replicate, repeat, and reproduce~\cite{renaud:2012}. To mitigate these issues especially in an academic setting with limited access to human user data, some studies suggested user simulation to analyze (interactive) information systems~\cite{maxwell:2015,maxwell:2016,camara:2022,zerhoudi:2022}. However, simulated users cannot yet compete with ``real'' human feedback. Recently, fully automatic evaluations, where the output of one system is judged by another, has been proposed as a possible way forward~\cite{liu:2023a,yue:2023}. But judging the output of generative models by means of other models has itself already been criticized~\cite{sakai:2023,bauer:2023,faggioli:2023}. 

\subsection{Steps of the Information Search Process}\label{steps-of-search-process}

To derive suitable evaluation objectives for generative ad hoc retrieval, we consider the general user side search process for which \citeauthor{vakkari:2016}~\cite{vakkari:2016} has suggested to differentiate four steps: search formulation, source selection, source interaction, and information synthesis (Figure~\ref{fig:evaluation-overview}, top row). Interestingly, each of these steps can be mapped to capabilities of generative retrieval systems.

First, during formulation, the user comes up with a specific query that expresses their information need. This is no different in generative retrieval systems, though what is called a `query' in~IR is often called a `prompt' for generative systems. To avoid confusion, we stick to the term `query'. Still, in this paper, we leave the formulation step entirely to the user who may iteratively adapt their search formulation. Yet, we do acknowledge that formulation may also be framed as a system task with the goal of enhancing the users' original query with more context or prompt templates, akin to query suggestion and query expansion in traditional retrieval.
Second, during selection, traditionally, the user is presented with a result list possibly containing surrogates like snippets that help to assess whether some result aligns with the user's information need and should be selected for further inspection. In generative retrieval, the selection step corresponds to the system selecting sources that contain potentially relevant information.
Third, during interaction, traditionally, the user analyzes the content of the selected results more deeply to extract and structure the relevant information that addresses the knowledge gap underlying the user's information need. In generative retrieval, this step also rather is on the system side by, for instance, attending the generation to previously retrieved pieces of information. 
Finally, during synthesis, traditionally, the user assembles the search outcome by combining relevant information from their interacted sources. In generative retrieval, synthesis corresponds to the model's inference and generation of the response text from the selected sources. Just like for human users, interaction and synthesis may commence concurrently.

\subsection{Evaluation Objectives}
\label{evaluation-objectives}
For each step of the search process, we define a corresponding generative retrieval-oriented evaluation objective. The objectives are not meant as evaluation steps, but rather as potential targets when evaluating a generative retrieval system as a whole.

\paragraph{Prompting Objective}

Corresponding to formulation are evaluation aspects related to a model's input prompt like preciseness (Does the prompt target the specific desired outcome?), ambiguity (Is the prompt unambiguous, targeting only the desired outcome?), or contextuality (Does the prompt provide sufficient context to delineate the information need?). While formulation is an important step to evaluate, it is out of the scope of our paper, as the formulation step in our setting is left to the user and as there already is extense work on prompt engineering~\cite{shin:2020,reynolds:2021,gao:2021,liu:2022a,sorensen:2022,white:2023,yang:2023}.

\paragraph{Retrieval Objective}

Corresponding to selection are evaluation aspects related to the retrieved sources from which a generative system draws its information. These sources (but also any relevant information that was not retrieved) directly impact the quality of the generated response. Therefore, the retrieval objective covers the assessment of a system's ability to identify relevant (aligning with the users' information need), diverse (covering a variety of information), informative (containing valuable information), and correct (providing accurate information) sources from a collection.

\paragraph{Grounding Objective}
Corresponding to interaction are evaluation aspects related to a generative retrieval model's ability to attend to source documents as evidence in response generation. Yet, such grounded text generation may suffer from confabulations~/~hallucinations of broadly two types~\cite{maynez:2020}: intrinsic confabulations (the model wrongly modifies information from the sources) and extrinsic confabulations (the model generates information not present in the sources). As both types can negatively impact the quality of a generated response~\cite{maynez:2020,lux:2020}, the grounding objective covers the assessment of a system's ability to correlate its generated output with information from source documents. This includes the ability to identify relevant information in the sources, to paraphrase information (restate some information correctly), and to establish consistency (not produce contradictions to other sources).

\paragraph{Presentation Objective}
Corresponding to synthesis are evaluation aspects related to a model's ability to condense relevant information from multiple sources into a single answer. Resembling multi-document summarization, the presentation objective covers the assessment of an answer's conciseness (at a level of granularity sensible given the topic and user~\cite{dang:2005}), coherence (uniform writing style in the answer), and accessibility (written in an understandable way; again, dependent on the user).

\subsection{Components of the User Model}
\label{components-of-user-model}

Developing a user model for generative~IR is challenging. The traditional user models were focusing on list~SERPs and might thus not apply to text~SERPs, as, for instance, the search process steps of selection and interaction are undertaken by the system instead. Additionally, little to no user behavior data on text~SERPs are available in the academic context (e.g., A/B~tests or laboratory studies) to base model validation or development on. To contribute a user model for generative~IR, we thus extrapolate from established evaluation practices in IR and related fields like question answering or summarization. We follow the considerations of \citet{carterette:2011}, who argues that a user model in~IR should include three distinct \mbox{(sub-)models}: 
\Ni
a utility model that induces a gain function by capturing how each result provides utility to a user, 
\Nii
a browsing model that induces a discount function by capturing how a user interacts with the results, and
\Niii 
an accumulation model that combines the individual gain and discount values by capturing how individual utility is aggregated.

\subsubsection{A Utility Model for Generative IR}

\begin{figure}[t]
    \centering
    \begin{tikzpicture}
    [
        grow                    = right,
        level 1/.style={sibling distance=2.5em, minimum width=5em, level distance=7em, edge from parent path = {(\tikzparentnode.east) -- ($(\tikzparentnode.east) + (1.5em, 0)$) |- (\tikzchildnode.west)}},
        level 2/.style={sibling distance=1em, text width=4em, level distance=8em, edge from parent path = {(\tikzparentnode.east) -- ($(\tikzparentnode.east) + (2em, 0)$) |- (\tikzchildnode.west)}},
        edge from parent/.style = {draw, line width = .5pt, {Triangle[open, width=2.1mm, length=2.1mm]}-},
        every node/.style       = {font=\small\sffamily,shape=rectangle},
        Label/.style = {font=\small\sffamily\bfseries, fill=white, text=black},
        Legend/.style = {text width = 5em, align = left}
    ]
\begin{scope}[local bounding box=bbox1]
    \node at (0,0) (anchor) [font=\sffamily\bfseries, minimum width=2em] {Utility}
    child {
        node (clarity) {Clarity}
        child {node (content) {Content}}
        child {node (language) {Language}}
    } 
    child {
        node (correctness) {Correctness}
        child {node (topical) {Topical}}
        child {node (factual) {Factual}}
    } 
    child {
        node (consistency) {Consistency}
        child {node (external) {External}}
        child {node (internal) {Internal}}
    }
    child {
        node (coverage) {Coverage}
        child {node (completeness) {Deep}}
        child {node (comprehensiveness) {Broad}}
    }
    child {
        node (coherence) {Coherence}
        child {node (logical) {Logical}}
        child {node (presentation) {Stylistic}} 
    };
    \draw[draw=black, line width=.5pt,decorate,decoration={brace,amplitude=.6em,mirror}] ($(internal.south east) + (0, 0.25em)$) -- node[right=.5em, text width=4em, align=center, anchor=west] {Response level} (presentation.north east); 
    \draw[draw=black, line width=.5pt,decorate,decoration={brace,amplitude=.6em,mirror}] (content.south east) -- node[right=.5em, text width=4em, align=center, anchor=west] {Statement level} ($(external.north east) - (0, 0.25em)$); 
    \coordinate (labelx) at ($ (anchor.east) + (9.05em,0) $);
    \draw[Blue5,fill=Blue2]     (clarity-|labelx)     circle (1ex);
    \draw[Yellow5,fill=Yellow2] (correctness-|labelx) circle (1ex);
    \draw[Red5,fill=Red2]       (consistency-|labelx) circle (1ex);
    \draw[Yellow5,fill=Yellow2] (coverage-|labelx)    circle (1ex);
    \draw[Blue5,fill=Blue2]     (coherence-|labelx)   circle (1ex);
\end{scope}

\begin{scope}[local bounding box=bbox1]
    \coordinate (legendx) at ($ (anchor.east) + (9em,-8.5em) $);
    \node (retrieval) [Legend, left=4em of legendx, anchor=east] {Retrieval};
    \node [left=0em of retrieval, anchor = east, fill=Yellow2, circle, draw=Yellow5] {};
    \node (grounding)  [Legend, right=0em of legendx, anchor = center] {Grounding};
    \node [left=0em of grounding, anchor = east, fill=Red2, circle, draw=Red5] {};
    \node (response) [Legend, right=5em of legendx, anchor = west] {Presentation};
    \node [left=0em of response, anchor = east, fill=Blue2, circle, draw=Blue5] {};
\end{scope}
\draw ($(bbox1.south west) - (1em,.25em)$) rectangle ($(bbox1.north east) + (1em,.5em)$);
\node at ($(bbox1.north) + (-.5em,.5em)$) [Label,anchor=center] {Evaluation objectives};
\end{tikzpicture}
    \caption{Taxonomy of utility dimensions in generative ad hoc retrieval; colors indicating the evaluation objectives.}
    \label{fig:utility-dimensions}
   \Description{A taxonomy tree growing from left to right which illustrates the utility dimensions in generative ad hoc retrieval. The root node is utility. It subdivides into two sub-levels, where each child on the first has two further childs on the third. Each child on the first level corresponds to an evaluation objective. From top to bottom, the nodes read: 1. coherence, subdivided into stylistic and logical (presentation objective); 2. coverage, subdivided into broad and deep coverage (retrieval objective); 3. consistency, subdivided into internal and external (grounding objective); 4. correctness, subdividing into factual and topical (retrieval objective); and 5. clarity, subdividing into language and content (presentation). Brackets to the right of tree indicate the level each dimension is measured at. Coherence, coverage, and consistency are measured on the response level. External consistency, correctness, and clarity are measured on the statement level.}
\end{figure}
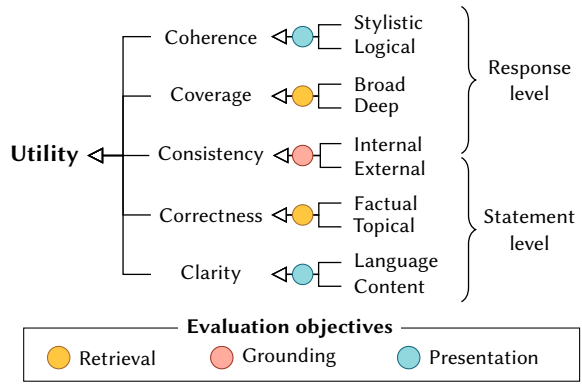

Surveying the literature on evaluation in~IR and in related fields, we identified ten utility dimensions applicable to generative ad hoc retrieval. Figure~\ref{fig:utility-dimensions} shows the dimensions grouped into five categories (coherence, coverage, consistency, correctness, and clarity) with color-coded corresponding evaluation objectives and  indicated granularity from which gain is obtained (statement level: from an individual statement in the response; response level: from the response as a whole).

\paragraph{Coherence}
Coherence is a response-level dimension of utility referring  to the presentation objective and involving the aspects of statement arrangement that should form a narrative without contradictions~\cite{radev:1998,shah:2021} (i.e., logical coherence: Is the response well-structured?) and of the writing style that should yield readable and engaging responses~\cite{jin:2020,capra:2023} (i.e., stylistic coherence: Does the response have a uniform style of speech?).

\paragraph{Coverage} 
Coverage is a response-level dimension of utility referring to the retrieval objective and measuring how well a user's information need is treated by the returned information; it can be subdivided into~\cite{cambazoglu:2021} broad coverage (i.e., whether the response covers diverse information~\cite{zheng:2012}), and deep coverage (i.e., whether the response provides in-depth and highly informative content~\cite{maxwell:2017}). 

\paragraph{Consistency}
Commonly observed problems with source-based text generation are inconsistencies between the sources and parts of the generated text~\cite{huang:2021} but also inconsistencies between the statements within a response. We refer to the first problem as external consistency, which is a statement-level dimension of utility involving the assessment of the consistency between a statement and its source document(s) to ensure that the generated text aligns in terms of content and context~\cite{maynez:2020,yue:2023,sakai:2023} (i.e., Is the statement accurately conveying from the sources?). External inconsistencies are often introduced through model confabulations~/~hallucinations~\cite{ji:2023} but they should be distinguished from factual correctness, as external consistency only assesses the alignment of a statement with the sources, and not with some objective truth. To the second consistency problem, we refer to as internal consistency, which is a response-level dimension of utility involving the assessment of the consistency between the responses' individual statements to ensure no contradictions~\cite{nishino:2019,sakai:2023,capra:2023}. It should be noted that this does not mean that different conflicting perspectives on a topic can not be reflected in the response, however, these should then be explained. Both notions of consistency refer to the grounding objective.

\paragraph{Correctness}
Correctness is a statement-level dimension of utility referring to the retrieval objective and measuring to which degree the information provided in the response is factual, reliable, and addressing the user's information needs. We subdivide correctness into factual and topical correctness. The former captures the degree to which a statement reproduces information that can be assumed as objectively true. Yet, outside of small-scale domain-specific evaluation studies~\cite{sallam:2023} fact-checking remains a challenge~\cite{nakov:2021} and is thus often reduced to a simpler approach, framing it in terms of verifiability~\cite{liu:2023a}, not truth. Here, the main requirement is that a piece of information can be attributed to a reliable reference~\cite{wikipedia:2023,yue:2023} (i.e., Does the statement state things that are verifiable?). Topical correctness captures whether a statement aligns with the user's information need~\cite{maddalena:2017,yang:2017,roitero:2018} (i.e., Does the statement state things within the scope of the user's information need?).

\paragraph{Clarity}
The response of a generative retrieval system should be expressed in a clear and understandable way~\cite{zhu:2009,sameki:2016}. This, on the one hand, comprises language clarity: concise~\cite{dang:2005,sakai:2023}, comprehensible~\cite{cambazoglu:2021}, lexically and grammatically correct, and user-accessible responses. Note that language clarity does not reflect fluency, which is assumed already at human-level for model-generated text~\cite{sakai:2023}, but rather the response being in the appropriate language register. For example, a technical query might warrant an academic style of writing in the response, while a joke question might afford a more jovial tone. On the other hand, clarity also comprises content clarity: in order to make a response explainable, the way a statement is written should always clearly communicate the most salient information~\cite{schuff:2022} and where it stems from~\cite{nourani:2019}. Both notions of clarity refer to the presentation objective at the statement level.

\subsubsection{A ``Browsing'' Model for Generative IR}

For list~SERPs, user interaction is modeled by a set-based or a ranking-based browsing model. In set-based browsing users are assumed to indiscriminately examine all retrieved documents (e.g., for systematic reviews), while in ranking-based browsing users are assumed to traverse the retrieved documents by rank, stopping when either their information need is fulfilled or the search is aborted~\cite{carterette:2011} (e.g., web search). Aborting a search is usually motivated by the information need being satisfied or the effort being too high to justify continuing to browse. Yet, in generative~IR, the selection and interaction steps of the search process are undertaken by the system, so that the user only has to read the (often short) generated text. This reduces the effect of effort-based stopping criteria, with most users only aborting their search when their knowledge gap is fulfilled or when the response is deemed insufficient. This is neither really set-based, as reading the response from the beginning and early stopping might occur, nor traditionally ranking-based, as aborting the search is not motivated by effort but rather by search (dis-)satisfaction.

Instead of a browsing model, we thus propose a \emph{reading} model for generative~IR to reflect the attention a user places on the response statements while reading. But as there are no dedicated studies on reading behavior for generative~IR yet, we turn to related work on reading behavior for document comprehension. In a literature survey, we identified six characteristics from which we deem three as appropriate for our reading model: progression~\cite{buscher:2012,li:2018,li:2019,zheng:2019,wu:2023} implies that users parse a document sequentially (i.e., reading the statements in their textual order), decay~\cite{frey:2013,li:2018,li:2019,zheng:2019,wu:2023} implies that the reading attention diminishes over the span of a text, and saturation~\cite{li:2018,li:2019} implies that users abort when they have read enough to satisfy their information need.

Besides these three characteristics, we deem three others as superfluous for our proposed reading model. First, perceived relevance may be heightened following a relevant statement~\cite{li:2018,zheng:2019} but we adopt the same restriction as static browsing models for ad hoc retrieval evaluation and neglect inter-statement effects~\cite{moffat:2013,moffat:2015}. Second, although reading attention may be highest around query terms~\cite{li:2018,zheng:2019}, our statement- and response-level utility granularities render per-token effects rather constant. Third, although users may skip non-relevant content during reading~\cite{li:2018,gwizdka:2014,buscher:2012}, we ignore this effect in the reading model as non-relevant statements will receive zero utility anyway.

Altogether, our proposed reading model thus reflects a sequential reading with decaying attention and early stopping when saturated. These properties can easily be related to the $C/W/L$~frame\-work~\cite{moffat:2017} of browsing models for list~SERPs. Sequential reading indicates that the frame\-work's assumption of a sequential process applies, decay (diminishing attention) is related to the framework's conditional continuation probability~$C$ and weight~$W$ (probability of a user reaching some step of a sequence), and early stopping (saturation) is related to the framework's probability~$L$ that indicates whether an item is the last one before aborting a search. Our proposed reading model can thus be operationalized as a monotonically decreasing weight function over statements that discounts the contribution of later statements in a response. At the same time, this directly induces a response structuring approach of putting the most important pieces of information first followed by less important details---similar to the inverted pyramid scheme of news articles~\citep{poettker:2003}.

\subsubsection{An Accumulation Model for Generative IR}

To combine gain and discount values over the statements of a response, we argue in favor of \emph{expected total utility} accumulation~\cite{carterette:2011,moffat:2013}. It considers the total utility a searcher accumulates from the whole response. Alternatively, measures could be based around estimating the total `cost'  of accruing information from the response in terms of the effort expended~\cite{carterette:2011}.  However, we argue that the effort is comparatively small in text~SERPs so that optimizing for it is not that suitable to reliably differentiate systems in evaluation.

\section{Operationalizing Evaluation}
\label{operationalizing-evaluation}

\begin{figure}[t]
\centering
\begin{tikzpicture}
[
    every node/.style = {font=\small\sffamily},
    Box/.style = {minimum height=3em, text width=6em, minimum width = 5em, align=center, rectangle, draw},
    Heading/.style = {font=\small\sffamily\scshape\bfseries, fill=white, text=black},
    Label/.style = {font=\footnotesize\sffamily, text=black},
    Arrow/.style = {-stealth, draw=Gray6, rounded corners=.25em,}
]
\coordinate (labelx) at (-13em,0);

\begin{scope}[local bounding box=bbox0]
\node[Box]                                       (systems)   {Generative re\-trieval systems}; 
\node[Box, dashed, left=1em of systems.west, anchor=east]  (topics)    {Document collection};
\node[Box, dashed, right=1em of systems.east, anchor=west] (documents) {Topic set w. queries};
\end{scope}
\path[draw=Gray5, line width=.8pt] (labelx |- bbox0.north west) -- (labelx |- bbox0.south west) node[Label, midway, above, rotate=90, minimum height=1.75em, inner sep=0] {Setup};

\begin{scope}[local bounding box=bbox1]
\node[Box, below=1em of systems.south, anchor=north] (segmentation) {Segmentation (optional)};
\end{scope}
\path[draw=Gray5, line width=.8pt] (labelx |- bbox1.north west) -- (labelx |- bbox1.south west) node[Label, midway, above, rotate=90,minimum height=1.75em, inner sep=0] {Statements};

\begin{scope}[local bounding box=bbox2]
\coordinate (assessment anchor) at ($(segmentation.south) - (0em, 1.5em)$);
\node[Box, left=2em of assessment anchor, anchor=north east] (initial assessment) {Reference-free \\ assessment};
\node[Box, right=2em of assessment anchor, anchor=north west] (repeated assessment) {Reference-based assessment};
\end{scope}
\draw[dashed] ($(bbox2.south west) - (2.25em, 1em)$) rectangle ($(bbox2.north east) + (2.25em, 1em)$);
\path[draw=Gray5, line width=.8pt] ($(labelx |- bbox2.north west) + (0em, 1em)$) -- ($(labelx |- bbox2.south west) - (0em, 1em)$) node[Label, midway, above, rotate=90, minimum height=1.75em, inner sep=0] {Utility assess.};

\begin{scope}[local bounding box=bbox3]
\node[Box, below = 1.5em of bbox2.south, anchor=north] (measure) {Evaluation measure};
\node[Box, dashed, left = 1em of measure.west, anchor=east] (user model) {User model};
\node[Box, right = 1em of measure.east, anchor=west] (ranking) {System ranking};
\end{scope}
\path[draw=Gray5, line width=1pt] (labelx |- bbox3.north west) -- (labelx |- bbox3.south west) node[Label, midway, above, rotate=90, minimum height=1.75em, inner sep=0] {Measure};

\path[Arrow] (topics.east) -- (systems.west) {};
\path[Arrow] (documents.west) -- (systems.east) {};
\path[Arrow] (systems.south) -- (segmentation.north) {};
\path[Arrow, dashed] (documents.south) -- ($(bbox2.north -| documents.south) + (0em, 1em)$) {};
\path[Arrow, dashed] (topics.south) -- ($(bbox2.north -| topics.south) + (0em, 1em)$) {};
\path[Arrow] 
    (segmentation.south) -- 
    ($(bbox2.north) + (0em,.5em)$) -- 
    ($(bbox2.north -| repeated assessment.north) + (0em,.5em)$) --
    (repeated assessment.north) {};
\path[Arrow] 
    (segmentation.south) -- 
    ($(bbox2.north) + (0em,.5em)$) -- 
    ($(bbox2.north -| initial assessment.north) + (0em,.5em)$) --
    (initial assessment.north) {};
\path[Arrow] 
    (repeated assessment.south) -- 
    ($(bbox2.south -| repeated assessment.south) - (0em,.5em)$) --
    ($(bbox2.south) - (0em,.5em)$) -- 
    (measure.north) {};
\path[Arrow] 
    (initial assessment.south) -- 
    ($(bbox2.south -| initial assessment.south) - (0em,.5em)$) --
    ($(bbox2.south) - (0em,.5em)$) -- 
    (measure.north) {};

\path[Arrow] (measure.east) -- (ranking.west) {};
\path[Arrow, dashed] (user model.east) -- (measure.west) {};
\path[Arrow, dashed] (user model.north) -- ($(bbox2.south -| user model.north) - (0em, 1em)$) {};
\path[Arrow, dashed] (initial assessment.east) -- (repeated assessment.west) {};
\end{tikzpicture}
\caption{Overview of the evaluation procedure for generative ad hoc retrieval. Given documents and topics, a system produces responses, which are segmented into statements, and assessed for utility, based on which an evaluation measure ranks systems by effectiveness. Solid lines indicate process flow, dashed lines contextual information sources.}
\Description{A flowchart illustrating the evaluation procedure for generative ad hoc IR. It subdivides the process into the four steps of setup, statement segmentation, utility assessment, and measuring utility. These steps have further graphical elements which correspond to the process described in the text of this section.}
\label{figure-evaluation-procedure}
\end{figure}
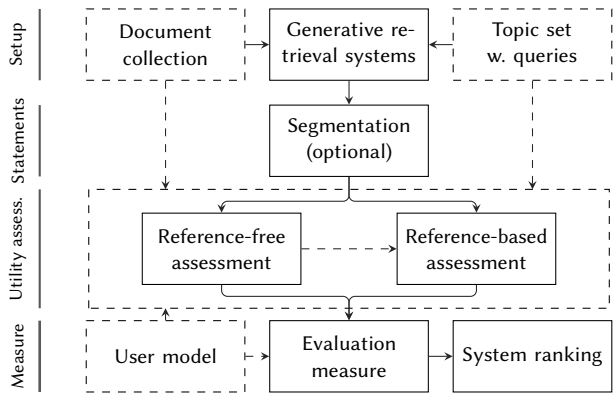

This section considers operationalizations of the proposed user model. The goal is to take stake in what possibilities exist for each step of the process, in an effort to illustrate the required components and how they can be implemented. These considerations are summarized in \autoref{figure-evaluation-procedure}, with each component (rows in the figure) described in a subsection below. The experimental setup encompasses a document collection, a set of topics reflecting the search task, and a set of generative retrieval systems to be evaluated (\autoref{data-sources}). Their responses to queries are (optionally) split into statements using a segmentation approach (\autoref{segmentation}). Statements are then assessed for their utility, distinguishing between assessment without prior reference, and assessment in relation to prior reference material (\autoref{utility-assessments}). Given annotations and an evaluation measure, the systems can then be ranked with respect to their effectiveness as indicated by an aggregated score (\autoref{effectiveness}). In each of these four steps, we survey relevant literature and juxtapose proposed evaluation processes with regard to their advantages and disadvantages in the context of the assumed user model.

\subsection{Experimental Setting}\label{data-sources}
The established approach for the reproducible evaluation of traditional retrieval systems in an academic context is offline evaluation~\cite{cleverdon:1967,sanderson:2010}. It encompasses a document collection, a set of topics reflecting the information needs stated by users, and the set of systems to be tested. Generative retrieval evaluation does not diverge from this basic procedure. Yet, the set of topics should include ones that reflect the search task for which generative retrieval systems are useful, i.e., the synthetic task posited in~\autoref{generative-search-task}. Furthermore, a ranking of documents could be pre-supplied for each topic's query in order to exclusively study the systems' synthesizing ability. These can be taken from a baseline retrieval system, shared task results~\cite{craswell:2020,craswell:2021,craswell:2022}, or query logs~\cite{reimer:2023}. While opting for offline evaluation allows to reuse established experiment infrastructure such as the TREC format specifications for run and utility judgment files,%
\footnote{\url{https://github.com/usnistgov/trec_eval/}}
generative retrieval systems introduce new requirements. Specifically, a run file represents a text~SERP, and should thus include the generated text instead of a ranked list of document identifiers. Utility judgments should be persisted together with the annotated text, since no static document identifiers are available.

\subsection{Segmenting Statements}\label{segmentation}

While the complete response provided by the system can be annotated as-is (this is especially warranted for response-level utility), in order to ease annotation, it can be segmented into retrieval units (suitable for statement-level utility). This approach of subdividing a response into smaller units is well established in evaluating generated texts in NLP~\cite{liu:2023b,nenkova:2007,dang:2007}, and has been proposed for IR as well~\cite{sakai:2011,sakai:2023}. Unit statements should be atomic, in the sense that an assessor should be able to make an informed and reliable decision about their utility with little to no surrounding context.

To this end, human judges can be employed to extract statements~\cite{dang:2006,dang:2007}, but the high effort and low repeatability, as well as the inability to assess the effectiveness of a new system without repeated human intervention renders this approach impractical in most settings. Automatic means of statement segmentation, comparable to the established task of web page segmentation~\cite{kiesel:2021}, could include splitting after each given reference (useful for experiments investigating grounding, as each statement has a clear attributable source), sentence-level splitting (useful for fine-grained utility dimensions such as correctness or coverage), or prompting the model to output already delineated statements.

\subsection{Assessing Utility}\label{utility-assessments}

Two different settings for collecting utility assessments can be discerned: 
\Ni
a direct assessment of the responses is carried out, without comparing to a separate ground truth; and
\Nii
the unjudged responses can be compared to pre-existing reference responses on the same document and/or query set. 
The first is similar to reference-free evaluation in summarization~\cite{fabbri:2021}, which instructs annotators to assess the summary directly, while the second is similar to reference-based evaluation in summarization~\cite{bhandari:2020}, which instructs annotators to assess the overlap between the system output and reference response, under the assumption that the reference response is the gold standard, or at least exemplary of utility. Not all utility dimensions can be judged on the generated text alone (as, e.g., clarity of language can), but also require information beyond the generated text (e.g., topical/factual correctness). We therefore discern reference responses and context: reference responses are one or more pre-existing texts to which a new response is compared, while context covers the assessment information required. An assessment made with context only is therefore deemed reference-free.

\paragraph{Reference-Free Assessment}

To operationalize reference-free evaluation for generative IR, the straightforward approach is to task human judges with assessing a given output. Yet, possibilities also include using the self-reported uncertainty of generative models with out-of-domain data~\cite{nalisnick:2019}, or relying on other generative models to assess the quality of the output, such as BARTScore~\cite{yuan:2021} or GPTScore~\cite{fu:2023}. Classifiers trained to estimate the magnitude of a utility dimension have also been used~\cite{kulesza:2004}. Ranking, either in a pairwise or listwise fashion is an additional form of assessment, i.e., tasking a judge with ordering statements of unknown utility with respect to a given utility dimension~\cite{gienapp:2020}, under the hypothesis that a response with higher utility will be ranked higher, too. 

\paragraph{Reference-Based Assessment}

To operationalize reference-based assessment, commonly a similarity measure is applied between reference and response. \citet{lazaridou:2022} evaluate their generative retrieval system for the task of question answering by matching words between generated response and the gold answer. Similarity, \citet{arabzadeh:2024a} assign relevance scores to candidate answers in a QA task by measuring their similarity to annotated ground truth data in latent space. Other content overlap metrics, though not necessarily transferable to the setup proposed here, such as BLEU~\cite{papineni:2002}, NIST~\cite{doddington:2002}, ROUGE~\cite{lin:2004} TER~\cite{snover:2006}, METEOR~\cite{banerjee:2005}, BERT Score~\cite{zhang:2020}, or MoverScore~\cite{zhao:2019} have been used to compare a generated text to a reference text, either in full or at the statement level. Ranking models have also proven useful for the relative assessment of generated texts in comparison to references, e.g., in machine translation~\cite{duh:2008,song:2011}, both in a listwise~\cite{li:2013} as well as a pairwise setting~\cite{guzman:2014,guzman:2015}. \citet{arabzadeh:2024a} implement a kind of pseudo-relevance feedback by retrieving candidate reference documents from a corpus, using highly-ranked ones as references.

\subsection{Measuring Effectiveness}
\label{effectiveness}

For statement-level evaluation, the individual utility of statements has to be combined into an overall score for the response. Effectiveness measures for the proposed aggregation model of expected total utility take the general form $\sum_{i=1}^k g(d_i)\cdot\sum_{j=i}^k p(j)$~\cite{carterette:2011}, where~$k$ is the evaluation depth, or in our case, response length, $g(d_i)$~is the utility of the statement at position~$i$, and~$p(j)$ is the probability of the user aborting their search immediately after position~$j$. The former is referred to as a gain function, given by the utility assessments of statements collected before. The latter as a discount function, chosen based on prior information about typical user behavior. The widely established measures of~DCG and nDCG~\cite{jarvelin:2002} used for traditional IR evaluation stem from this family of measures~\cite{carterette:2011} and seem suitable for generative retrieval evaluation as well. Yet, they assume a logarithmic discount function. It is currently unclear if this is an appropriate choice to model the effect of decay and saturation in the proposed reading model for generative IR. While the family of measures is thus applicable, the concrete choice of measure needs further empirical validation from user experiments.

For response-level evaluation, two choices for measuring effectiveness exist: either utility is annotated directly for a response, or it is aggregated from individual statement utility. While the latter seems counterintuitive to the response-level vs. statement-level distinction made for utility before, note that the level of granularity on which a  utility dimension is defined, and the level of granularity at which annotations are collected can differ. Response-level utility may be aggregated from annotations of individual statements, or statement utility may be derived from annotations of the whole response. For example, consider the response-level utility dimension of broad coverage. It can be estimated by measuring the breadth of topics occurring over all statements, hereby annotating which topics occur in each statement. The previously motivated family of DCG-type measures can be extended to support such evaluation. For example, measure modifications similar to $\alpha$-nDCG~\cite{clarke:2008} that reward a diverse set of topics in a ranked list can be made for generative IR as well. Independent of how a single score is produced for each response, the final system score is aggregated over multiple topics, increasing  robustness and enabling statistical testing.

\subsection{Comparison with Existing Frameworks}

Two other approaches for the evaluation of generative retrieval systems have been proposed recently: SWAN~\cite{sakai:2023} and  EXAM~\cite{sander:2021}. The starting point of both is a text~SERP response, albeit less formalized and without considering the synthetic search task it enables. 

SWAN follows a similar approach as is proposed here, first establishing the notion of `information nuggets', i.e., statements, that constitute the response. Then, a total of 20~categories are described, indicating how a nugget may be scored. The individual nugget scores are then averaged over the whole response. Here, too, two different levels of score categories, i.e., utility dimensions are considered. While similar, our approach and SWAN differ in three important aspects. First, we base our method on a theoretical foundation in the form of a user model, whereas SWAN is mainly motivated from a standpoint of practicability. Second, SWAN is geared towards conversational search, while we consider the ad hoc search task. And third, the utility dimensions we propose differ from SWAN due to the shift in scope: We exclude dimensions specific to conversational search (e.g., recoverability, engagingness), and also those which do not serve to operationalize evaluation for the synthetic search task specifically (such as non-toxicity, robustness to input variations, etc.). The majority of the remaining utility dimensions from SWAN can be mapped to ours.

EXAM takes a completely different approach. Instead of directly evaluating inherent qualities of the generated text, it considers the downstream effectiveness of a Q\&A system that ingests the generated answer on multiple-choice questions. The hypothesis is that the correctness of its responses are correlated with the quality of the generated text it uses as input. Being an automatic evaluation method, this allows for rapid experimentation, yet exhibits three major drawbacks: It offers no fine-grained insight into the quality of the generated text, it is not grounded in a user model, and it requires a suitable Q\&A~system, impacting reliability and comparability, since there are no accepted standards.

In sum, our approach can be related to existing methods in terms of compatibility, complementarity, and consistency. Our approach is compatible with SWAN, as it is derived from similar assumptions, yet adding a theoretical foundation, and constructed with a different search task in mind. Our approach is complementary to EXAM, as our focus is on fine-grained, reliable, user-oriented evaluation, whereas EXAM excels for rapid, system-oriented experimentation with little overhead. Furthermore, our approach is consistent with traditional IR~evaluation techniques, making only small adaptations to the utility, browsing, and aggregation models to accommodate the new search paradigm. We believe that this renders much of the work on methods and theoretical foundation for traditional IR evaluation still applicable.

\section{Conclusion}
\label{conclusion}

Generative retrieval introduces a new paradigm for the retrieval of information. With it comes the need to measure and understand new utility dimensions that make text~SERP responses from generative retrieval systems relevant to a user's information need. In this paper, we have extrapolated a theoretical foundation for the evaluation of generative retrieval systems from traditional IR and related disciplines. First, we established that the search task of generative ad hoc retrieval goes beyond acquiring information, and instead enables the condensation of information, a process we dub the `synthetic search task'. Second, we proposed a new user model that accommodates this task, including evaluation objectives, utility dimensions, and a browsing model for text SERPs. Finally, we outlined how one could operationalize the evaluation of generative retrieval systems, surveying how existing evaluation approaches relate to, and could fit into the proposed methodology.

Many techniques for constructing generative retrieval systems are currently emerging, but evaluating their output is still a non-standardized and thus hardly comparable effort, lacking a theoretical motivation. We have provided our vision of a comprehensive approach for evaluating generative retrieval systems. Yet, we believe that user experiments are needed to effectively apply this theoretical motivation, and studying its reliability and validity. This requires a meta-evaluation, such as recently started by~\citet{arabzadeh:2024b}, of both, existing measures and measures modified for generative IR specifically, to study how well they align with user preferences, and to study the proposed utility dimensions and their ability to reflect user satisfaction, similar to studies conducted for traditional IR~\cite{cambazoglu:2021}. In addition, investigating user interactions with generative retrieval systems is warranted; for example, are user clicks on cited documents in a generated response indicative of their relevance or the user's disbelief, or will generative retrieval make clicks superfluous?

\paragraph{Limitations}
The evaluation process we propose in this paper is limited in two ways. First, we opted for a \emph{holistic} evaluation of text~SERPs, i.e., instead of evaluating the pipeline of components that constitute the generative retrieval system individually, we focus on evaluating the final response. Second, the evaluation is additionally limited to answer the question if a generative retrieval system is successful at supporting the synthetic search task. This does not consider the more general evaluation objectives that all search systems are subject to (such as bias, fairness, ethicality, or user privacy). In that sense, our considerations are \emph{specific} to generative IR, disregarding the evaluation of \emph{systemic} aspects of IR as a whole. This is not meant to deemphasize the importance of evaluating, e.g., bias in search results, but rather considers it to be outside the scope of this paper.

\begin{acks}
This publication has been partially supported by the ScaDS.AI Center for Scalable Data Analytics and Artificial Intelligence, funded by the Federal Ministry of Education and Research of Germany and by the S{\"a}chsische Staatsministerium f{\"u}r Wissenschaft, Kultur und Tourismus; by a Research Fellowship for Harrisen Scells from the Alexander von Humboldt Foundation; and by the OpenWebSearch.eu project, funded by the European Union (GA~101070014). 
\end{acks}

\bibliographystyle{ACM-Reference-Format}
\balance
\bibliography{sigir24-generative-ir-evaluation-lit}

\end{document}